\documentclass[aps,prl,twocolumn,groupedaddress,showkeys]{revtex4}
\usepackage{amsmath}
\usepackage{amssymb}
\usepackage{graphicx}
\usepackage{xcolor}

\begin{document}

\title{Harnessing exciton-exciton annihilation in two-dimensional semiconductors} 

\author{Eric Linardy,$^{1,2}$ Dinesh Yadav,$^{3,4}$ Daniele Vella,$^{1,2}$ Ivan A. Verzhbitskiy,$^{1,2}$\\
Kenji Watanabe,$^{5}$ Takashi Taniguchi,$^{5}$ Fabian Pauly,$^{3,4}$
Maxim Trushin,$^{2*}$ and Goki Eda$^{1,2,6*}$}

\affiliation{$^1$Department of Physics, National University of Singapore, Singapore}
\affiliation{$^2$Centre for Advanced 2D Materials, National University of Singapore, Singapore}
\affiliation{$^3$Okinawa Institute of Science and Technology Graduate University, Onna-son, Okinawa 904-0495, Japan}
\affiliation{$^4$Department of Physics, University of Konstanz, 78457 Konstanz, Germany}
\affiliation{$^5$National Institute for Material Science, 1-1 Namiki, Tsukuba 305-0044, Japan}
\affiliation{$^6$Department of Chemistry, National University of Singapore, Singapore}


\date{\today}

\begin{abstract}
Strong many-body interactions in two-dimensional
(2D) semiconductors give rise to efficient exciton--exciton
annihilation (EEA). This process is expected to result in the
generation of unbound high energy carriers. Here, we report an
unconventional photoresponse of van der Waals heterostructure
devices resulting from efficient EEA. Our heterostructures, which
consist of monolayer transition metal dichalcogenide (TMD),
hexagonal boron nitride (hBN), and few-layer graphene, exhibit
photocurrent when photoexcited carriers possess sufficient energy
to overcome the high energy barrier of hBN. Interestingly, we find
that the device exhibits moderate photocurrent quantum efficiency
even when the semiconducting TMD layer is excited at its ground
exciton resonance despite the high exciton binding energy and large transport barrier. Using ab initio calculations, we show that EEA
yields highly energetic electrons and holes with unevenly distributed energies depending on the scattering condition. Our findings
highlight the dominant role of EEA in determining the photoresponse of 2D semiconductor optoelectronic devices.
\end{abstract}

\maketitle

\begin{widetext}
\hfill
\begin{figure}
 \includegraphics[width=0.75\textwidth]{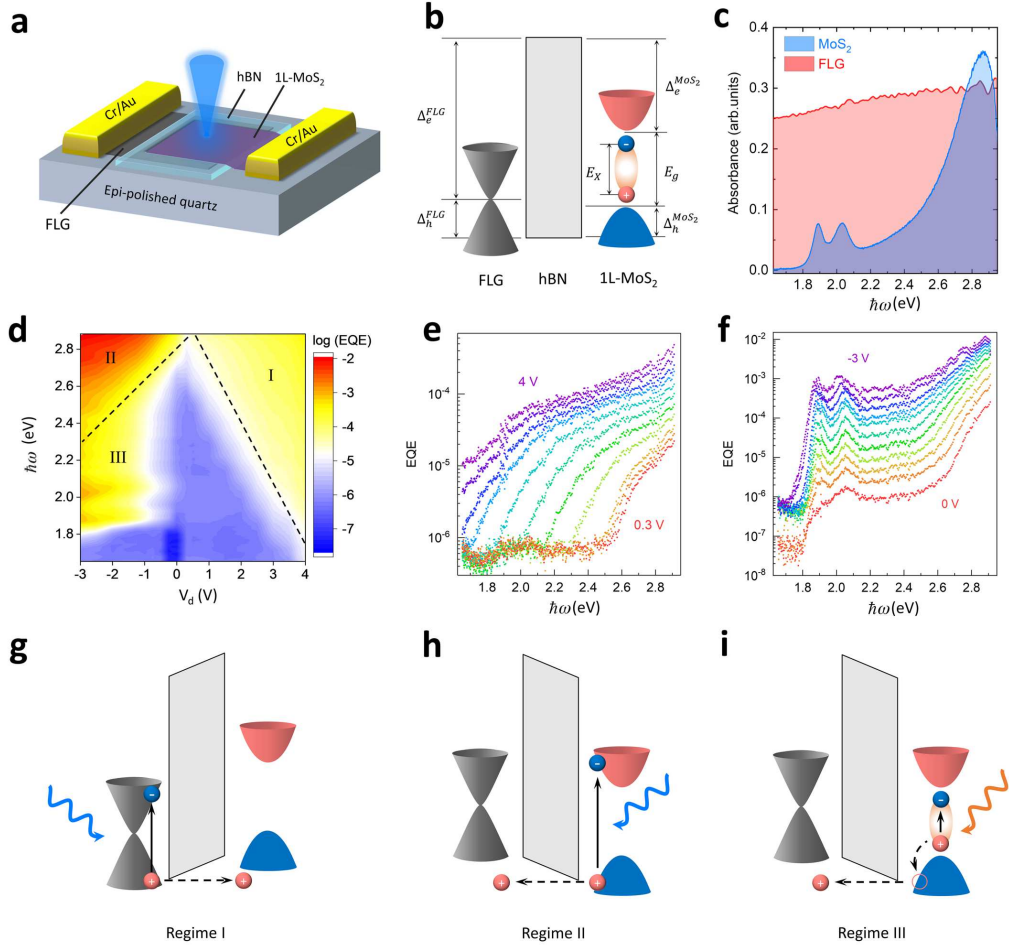}
 \caption{Photocurrent response of FLG/hBN/MoS$_2$ heterostructure device. 
 {\bf a}, Schematic of a FLG/hBN/MoS$_2$ heterostructure irradiated by a laser beam.
 {\bf b}, Schematic band diagram in flat band condition. $E_X$ and $E_g$ represent excitonic and quasiparticle 
 band gap of monolayer MoS$_2$. $\Delta_h^\mathrm{FLG}$, $\Delta_h^\mathrm{MoS_2}$, $\Delta_e^\mathrm{FLG}$,
 and $\Delta_e^\mathrm{MoS_2}$
represent the energy barrier for interlayer hole and electron transport at FLG/hBN and hBN/MoS$_2$ interfaces, respectively.
 {\bf c}, Absorbance spectrum of FLG and monolayer MoS$_2$.
{\bf d}, Color plot of EQE measured as a function of the excitation photon energy and bias voltage.
The blue region corresponds to low EQE below the detection limit. The origin of finite EQE in the yellow and red regions I, II, and III is explained in panels  {\bf g},  {\bf h}, and  {\bf i}, respectively. 
Dashed lines are guides to the eye, separating the regimes I, II, and III.
 {\bf e}, EQE measured as a function of the excitation energy at different forward biases 
 (0.3, 0.4, 0.5, 1, 1.5, 2, 2.5, 3, 3.5, 3.75, 4 V).
  {\bf f}, EQE measured as a function of the excitation energy at different reverse biases
  (0, --0.25, --0.5, --0.75, --1, --1.25, --1.5, --1.75, --2, --2.5, --3 V). 
   {\bf (g--i)}, Schematic describing the photocurrent generation mechanism. 
   In {\bf g}, hole is photoexcited in FLG and transferred to MoS$_2$ across the hBN barrier in forward bias.
   In {\bf h}, photoexcited high energy non-thermalized hole is emitted from MoS$_2$ to FLG in reverse bias.
   {\bf i}, Hot hole is created following formation of ground exciton in MoS$_2$ and transferred to FLG.
   Here, the excitation energy is below the quasiparticle band gap of MoS$_2$ and the heterostructure is under reverse bias.
 }
 \label{f1}
\end{figure}
\end{widetext}

Many-body effects manifest themselves as unique photoresponse in nanoscale materials \cite{mak2016photonicsreview}.
Exciton--exciton annihilation (EEA), a four-body interaction involving the energy and momentum transfer between two holes and two electrons, is known to be highly efficient in nanostructures with reduced dimensions, such as quantum dots \cite{qd2013,qd2016}, graphene nanoribbons \cite{nanoribbonsXXA2016}, 
carbon nanotubes \cite{NanotubesXXA2005,NanotubesXXA2010},
and polymer chains \cite{1DorganicXXA2006,1DorganicXXA2000}.
Recent studies \cite{XXAobservation2014-1,XXAobservation2014-2,XXAobservation2014-3}
indicate that the EEA rate in two-dimensional (2D) Group VI transition metal
dichalcogenide (TMD) semiconductors is substantially enhanced even when compared to their bilayer and trilayer samples \cite{fewlayerWS2XXA2015}. While the EEA in 2D semiconductors leads to undesirable efficiency droop in photoluminescence (PL) \cite{Science2015javey}  and electroluminescence (EL) \cite{NanoLett2017goki},
energy conservation requires hot electron-hole pairs to be generated upon annihilation of individual excitons. 
Recent study showed that the EEA-generated hot carriers can be harnessed for remarkably efficient
optical upconversion in 2D TMD semiconductors \cite{upconversion2018,upconversion2017},
and their heterostructures \cite{upconversion2019}, inspiring prospects for harnessing
highly efficient EEA process in 2D materials to generate hot
carriers for energy harvesting application.

Herein, we report photoresponse originating from EEA-generated hot carriers in a metal-insulator-semiconductor
(MIS)-type van der Waals heterostructure device. In this study, ``hot" carrier refers to non-thermalized high energy carriers
generated by EEA in contrast to thermalized hot carriers in 
graphene-based photothermoelectric devices \cite{gabor2011hot,song2011hot,sun2012ultrafast}.
The MIS-type heterostructure consists of few-layer graphene (FLG),
hexagonal boron nitride (hBN), and monolayer molybdenum disulfide (MoS$_2$) (Figure \ref{f1}a).
The bright optical microscope and atomic force microscope (AFM) image of one of the MIS
devices is displayed in the Supporting Information Figure S1.
Figure \ref{f1}b depicts the schematic band diagram of the heterostructure.
Because of the low electron affinity and wide band gap of hBN (Figure \ref{f1}b),
it acts as an insulating barrier which blocks charge transfer between the FLG and the MoS$_2$ layer.
However, the band alignment between hBN with FLG and MoS$_2$ reveals significantly lower energy barrier for
interlayer hole transfer compared to electron transfer \cite{NanoLett2017vu,britnell2012field,ma2016tuning,choi2013controlled,jeong2016metal}.
Hence, hBN acts as a carrier- and energy-selective tunnel barrier that selectively favors hot hole transfer.
Under forward bias above a specific threshold voltage ($V_\mathrm{d}>V_\mathrm{th}$), the device
exhibits EL at low threshold current densities, evidencing hole tunneling from FLG to MoS$_2$.
The tunnel barrier for holes in reverse bias ($V_\mathrm{d}<0$) is similarly low, however, the device
remains highly insulating due to the absence of holes in MoS$_2$. The charge transport and EL behavior of the tunnel diode are
discussed in Supporting Information S3 and in a previous study \cite{NanoLett2017goki}.

We investigate the photocarrier dynamics of the heterostructure
by measuring the bias-dependent spectral features in the photocurrent generated by photons of varying energy
ranging from 1.65 to 2.91 eV. All photocurrent measurements were performed with lock-in technique 
(Supporting Information S1) \cite{mechPhotoMoS2} at room temperature and in vacuum ($10^{-5}$ mbar).
On the basis of the band alignment of the heterostructure,
non-negligible photocurrent is only expected when the
photoexcited carriers gain sufficient energy to overcome the
potential barrier due to hBN (Figure \ref{f1}b). Note that the FLG
exhibits relatively uniform absorption across the visible
frequencies whereas monolayer MoS$_2$ absorption exhibits
prominent excitonic resonances with three excitonic peaks (Figure \ref{f1}c).
Two types of excitons, A and B, at 1.89 eV and 2.03 eV arise from spin--orbit split bands. 
Figure \ref{f1}d shows the external quantum efficiency (EQE),
the incident photon to converted electron ratio, 
as a function of the excitation energy $\hbar\omega$ and bias $V_\mathrm{d}$. 
The EQE value is calculated from the measured photocurrent ($I_\mathrm{PC}$) as EQE=$(I_\mathrm{PC}/P)\cdot(\hbar\omega/e)$,
where $P$ is the optical power, and $e$ is the elementary charge.

The spectral features are distinctly different in the forward (Figure \ref{f1}e) and reverse (Figure \ref{f1}f) bias regimes, but the common trend is the presence of distinct threshold behaviour. In forward bias, the photocurrent remains negligible at low photon energy before increasing exponentially above the threshold photon energy $\hbar\omega_\mathrm{th}$.
The $\hbar\omega_\mathrm{th}$, in turn, decreases linearly with increasing $V_\mathrm{d}$, as shown by the dashed line in Figure \ref{f1}d. The photocurrent spectrum is featureless above the threshold with no traces of excitonic absorption, suggesting that photocarriers generated in FLG play a dominant role in photocurrent generation (regime I). The maximum $\hbar\omega_\mathrm{th}$ is $\sim 2.6$ eV at low bias. Here, $\hbar\omega_\mathrm{th}/2$
coincides with the difference between the neutrality point of FLG and the valence-band maximum of hBN, corresponding to the
energy barrier for interlayer hole transfer from FLG to MoS$_2$ based on previous studies on photocarrier transport across
graphene/hBN interface \cite{britnell2012field,ma2016tuning} (Figure \ref{f1}b).
This indicates that the interlayer charge transport of non-thermalized photo-carriers,
possibly by Fowler--Nordheim tunnelling or over-barrier direct transport, is responsible for the observed
photocurrent above $\hbar\omega_\mathrm{th}$ (Figure \ref{f1}g).
Under reverse bias, EQE exhibits a similar threshold behaviour but with two prominent 
peaks corresponding to A and B excitonic absorption resonances. 
At the first sight, the threshold behavior can be explained as the onset of hot hole direct tunneling from MoS$_2$ to FLG (see Figure \ref{f1}h, regime II), similar to the forward bias case albeit in opposite direction. On the other hand, the relatively strong photocurrent in the low-energy excitation regime is unexpected due to high binding energy of excitons and their insufficient kinetic energy to overcome the potential barrier of hBN. Remarkably, the EQE at the exciton resonance in this regime III is comparable to those in regime II, specially under strong reverse bias. 
The observation of a finite photocurrent at the excitation energy below the quasiparticle
bandgap of MoS$_2$ indicates exciton dissociation generating hot holes with sufficient excess energy to overcome potential
barrier due to hBN (see Figure \ref{f1}i). 

\begin{figure}
\begin{center}
 \includegraphics[width=\columnwidth]{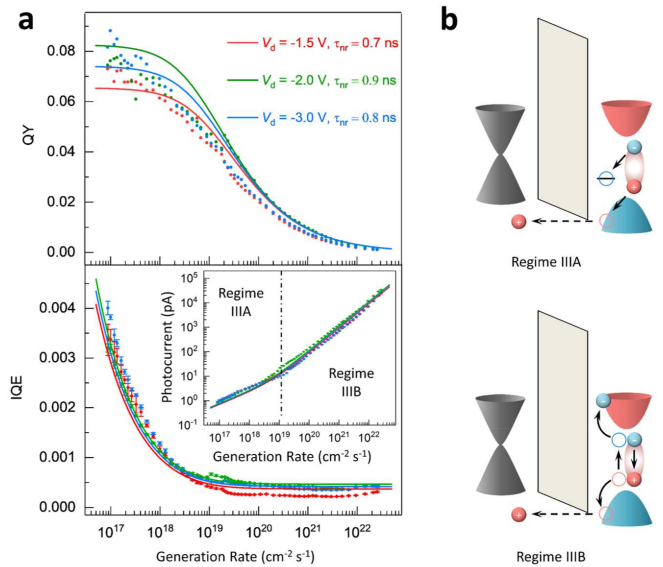} 
\end{center}
 \caption{Photocurrent and PL measurement of exciton--exciton annihilation in monolayer MoS$_2$.
 {\bf a},  QY and IQE  measured under the same experimental conditions
 and consistently fitted as functions of $g_\mathrm{eh}$ 
 using different values of $\tau_\mathrm{nr}$ for each $V_\mathrm{d}$. The experimental data is taken at voltages
lower than $-$1 V to reduce electron doping and to obtain higher QY \cite{Science2019javey}.
 QY is deduced from equation (\ref{rate1}) with $k_\mathrm{XX}=0.1$ cm$^2$/s and 
 $\tau_\mathrm{r}=10$ ns.
 IQE is derived from equation (\ref{rate2}) assuming $\tau_\mathrm{tr}=3.75$ ns and $k_\mathrm{CC}=5\times 10^{4}$ cm$^2$/s.
 The inset shows the photocurrent measured and calculated from equation (\ref{rate2}).
 The trend changes at a generation rate of about $10^{19}$ cm$^{-2}$s$^{-1}$,
 indicating activation of EEA processes. The excitation energy is about 1.96 eV. 
 {\bf b}, Schematics describing photocurrent generation mechanisms. The upper panel shows a charge-trapping process generating a high-energy free hole corresponding to regime IIIA in {\bf a}.
 The lower panel describes an EEA process generating a pair of hot carriers, corresponding to regime IIIB in
 {\bf b}.}
 \label{f2}
\end{figure}

There are two possible mechanisms for hot holes generation
through exciton dissociation: disorder-assisted exciton decay
and EEA \cite{Moody16}.
The exciton dissociation rate through the first
mechanism varies linearly with exciton density \cite{NanoLett2015ehrecomb} while the
latter varies quadratically \cite{XXAparameters}.
To distinguish these two mechanisms, we measured photoluminescence (PL) and the
photocurrent as a function of excitation power at the excitation
energy near the ground-state exciton resonance, well below the
quasiparticle band gap of MoS$_2$ and energy required for hole tunnelling from monolayer MoS$_2$ to FLG (Figure \ref{f1}d,f).
This allows us to determine whether the same rate constants
consistently describe the behavior of these interdependent phenomena.
Figure \ref{f2}a shows the power dependence of the PL quantum yield (QY) 
and photocurrent internal quantum efficiency (IQE) measured concurrently on the same sample with 633 nm (1.96 eV) laser as the excitation source. 
The experimental details and the calculation of PL QY and IQE are
described in Supporting Information S1.

The PL QY is nearly constant at low photocarrier generation rates and gradually decreases with increasing radiation power.
In contrast, the photocurrent IQE drops initially and saturates to a finite value in the high-power regime.
The steady-state exciton concentration $n_\mathrm{X}$ can be found by balancing the electron-hole generation rate $g_\mathrm{eh}$ with the total exciton decay rate as
\begin{equation}
 \label{rate1}
 k_\mathrm{XX} n_\mathrm{X}^2 + \frac{n_X}{\tau_\mathrm{X}} = g_\mathrm{eh},
\end{equation}
where $k_\mathrm{XX}$ is the EEA rate, $\tau_\mathrm{X}^{-1} = \tau_\mathrm{r}^{-1}  + \tau_\mathrm{nr}^{-1}$ is the reciprocal exciton lifetime with $\tau_\mathrm{r}$ and $\tau_\mathrm{nr}$ being the radiative and non-radiative exciton relaxation times, respectively.
The latter is mostly due to charge traps \cite{NanoLett2015ehrecomb} 
with an additional contribution from electron-exciton annihilation \cite{Science2019javey},
which is particularly relevant for n-doped samples.
The charge traps are expected to be sulfur vacancies, which are the most energetically favorable
type of defect \cite{sulfur_vac2014}.
By tuning the Fermi level through an external electric field, we change the equilibrium
occupation of traps altering their charge trapping efficiency.
Thus, the electron-exciton annihilation rate also changes with bias \cite{Science2019javey}.
Hence, $\tau_\mathrm{nr}$ differs for each PL QY curve shown in Figure \ref{f2}a.
The theoretical QY is given by the ratio between radiative recombination and
generation rates,
$\mathrm{QY} = n_\mathrm{X}/(\tau_\mathrm{r} g_\mathrm{eh})$,
where $n_\mathrm{X}$ is the solution of equation (\ref{rate1}).
In the low-power  (or linear-response) limit, the QY approaches a constant 
given by the ratio $\tau_\mathrm{X}/\tau_\mathrm{r}$ that at the same time determines the maximum QY for a given voltage.
In the high-power limit the QY decreases with increasing $g_\mathrm{eh}$ as
$\mathrm{QY} \sim 1/(\tau_\mathrm{r} \sqrt{g_\mathrm{eh}k_\mathrm{XX}})$ because EEA reduces the steady-state concentration of excitons. The kinetic model in equation 1 describes the power-dependence of QY observed in experimental data with fitting parameters of $k_\mathrm{XX}=0.1$ cm$^2$/s, $\tau_\mathrm{r}=10$ ns and $\tau_\mathrm{nr} = 0.7$ to $0.9$ ns. The fitting parameters are consistent with earlier studies in monolayer MoS$_2$ \cite{XXAobservation2014-1, XXAparameters} and other 2D semiconductors such as  MoSe$_2$\cite{XXAobservation2014-3}, WS$_2$ \cite{fewlayerWS2XXA2015}, and WSe$_2$\cite{XXAobservation2014-2}.




While EEA leads to reduction of exciton population and therefore diminishes the PL QY,
it generates hot carriers that can overcome the hBN barrier and contribute to photocurrent. For the IQE model, we consider the concentration of such hot holes denoted by $N$.
Here, EEA competes with high-energy carrier-carrier collisions leading in particular to
impact ionization, which is described by the rate $k_\mathrm{CC}$ of the order of $10^4$ cm$^2$/s.
The same process makes the high-energy holes thermalize by collisions with lower energy carriers, hence,
lose the energy required to escape the semiconductor.
At the typical $N\sim 10^9$ -- $10^{10}$ cm$^{-2}$,
the corresponding carrier-carrier collision time is of the order of $10$ -- $100$ fs,
consistent with literature values in bulk \cite{themalizationGaAsInP} and 2D semiconductors
 \cite{thermalizationMLMoS2,thermalizationFLMoS2}.
The rate equation for $N$ can be written as
\begin{equation}
 \label{rate2}
 k_\mathrm{XX} \left(g_\mathrm{eh}\tau_\mathrm{X}\right)^2 +
 \frac{g_\mathrm{eh}\tau_\mathrm{X}}{\tau_\mathrm{nr}} = k_\mathrm{CC} N^2 + \frac{N}{\tau_\mathrm{tr}},
\end{equation} 
where $\tau_\mathrm{tr}$ is the interlayer transport time (ns scaled),
and the corresponding $\mathrm{IQE} = N/(\tau_\mathrm{tr} g_\mathrm{eh})$
is given by the ratio between the transport and generation rates.
Equation (\ref{rate2}) is deduced in Supporting Information S1.
In contrast to a recent work \cite{ehliquid2019}, we do not use ultrafast pulses
and consider the rate equation in the steady-state limit. The model does not account for the
carrier-phonon interactions because photocarrier thermalization is dominated by the carrier-carrier scattering away
from the band edges \cite{ciccarino2018dynamics}.

The defect-assisted non-radiative recombination 
is also responsible for the hole generation
(as shown in Figure \ref{f2}b) 
that connects equations (\ref{rate1}) and (\ref{rate2}).
In the simplest case, when no barrier exists and
all the photoexcited carriers can escape the semiconductor, 
the terms containing $k_\mathrm{XX}$ and $k_\mathrm{CC}$ are not relevant,
and we obtain $\mathrm{IQE} = \tau_\mathrm{X}/\tau_\mathrm{nr}$.
In MIS heterostructure case, the hBN barrier places the hole escape window to rather high energies 
($\sim 0.4$ eV or greater with respect to the top of the MoS$_2$ valence band),
and the holes rapidly leave the window while thermalizing.
Because of carrier-carrier scattering being a quadratic function of carrier concentration,
the resulting IQE is strongly diminished with increasing excitation rate. However, EEA compensates the carrier-thermalization process at high excitation
densities and prevents the IQE from falling to zero. 
In the limit of high photocarrier generation rates, we obtain a saturated IQE independent of $g_\mathrm{eh}$ given by
\begin{equation}
\label{IQEsaturate}
\mathrm{IQE}_\mathrm{sat} = \frac{\tau_\mathrm{X}}{\tau_\mathrm{tr}}\sqrt{\frac{k_\mathrm{XX}}{k_\mathrm{CC}}},\quad 
k_\mathrm{XX} g_\mathrm{eh} \tau_\mathrm{X} \tau_\mathrm{tr} \gg 1.
\end{equation}
This model fits the experimentally observed trend, as shown in Figure \ref{f2}a.
The saturation of the IQE at high photocarrier generation rates is a clear manifestation 
that the photocurrent generation involves EEA.

\begin{figure}
\begin{center}
 \includegraphics[width=\columnwidth]{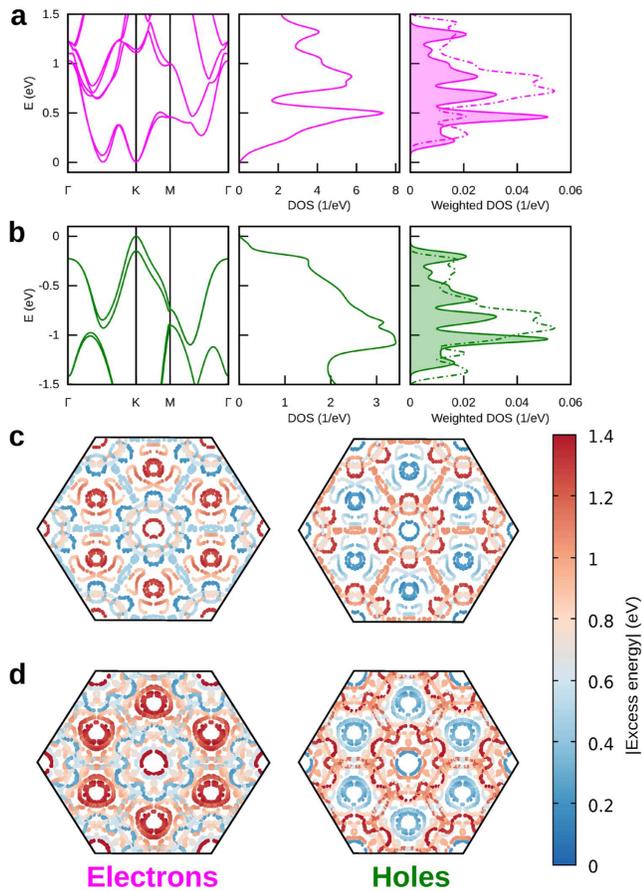}
\end{center} 
 \caption{Energy of EEA-derived carrriers in MoS$_2$.
 {\bf a}, Conduction band structure (left), conduction band DOS (center), and the DOS weighted
by the EEA probability for the type-A (filled solid curve) and type-B (dot-dashed curve) excitons (right). 
 {\bf b}, Same as {\bf a} for valence bands.
 {\bf c}, The Brillouin-zone regions accessible to the electrons (left) and holes (right), derived for the type-A exciton EEA. The regions are color-coded with the electron and hole excess energies counted from the valence- and conduction-band edges, respectively. {\bf d}, Similar to {\bf c} for the type-B excitons.}
 \label{f3}
\end{figure}

In EEA process, nonradiative exciton recombination ionizes
another exciton. Considering energy conservation law, the
ionization energy is equal to the energy released
during recombination process. This recombination energy is
equal to excitation energy $E^\mathrm{A}_\mathrm{X}$
and $E^\mathrm{B}_\mathrm{X}$ for excitons A and B, respectively.
Ionization energy equals the sum of the exciton binding energy and the excess energy shared between the free
electron and hole created.
The exciton binding energies are given as follows:
$E_\mathrm{b}^{\mathrm{A},\mathrm{B}}=E_\mathrm{g}-E^{\mathrm{A},\mathrm{B}}_\mathrm{X}$. 
Further, the momentum
conservation law dictates that the electron and hole wave
vectors of ionized exciton must be opposite in direction,
although equal in magnitude, in the absence of Umklapp
processes. Hence, there is a finite range of possible allowed
states in electronic structure of MoS$_2$ that satisfies the energy
and momentum restrictions. Here, we identify the allowed
transition states within the first Brillouin zone of MoS$_2$ EEA
processes to generate hot carriers that can contribute to
photocurrent generation using density functional theory
(DFT) modeling. The wave vectors are counted with respect
to the corner of the Brillouin zone, where the electron-hole
pairs emerge due to EEA.
Through scissor operators, we manually adjust the quasiparticle bandgap to the experimental value of $E_\mathrm{g}=2.24$ eV
since the theory does not reliably predict quasiparticle energies \cite{martin2004electronic}.
Figure \ref{f3}a,b shows the band structure, density of states (DOS), and 
the DOS weighted with the probability to occupy the states due to EEA.
Figure \ref{f3}c,d shows the final wave vectors of electrons and holes 
in the Brillouin zone with their excess energies encoded in color.
Note that the final electron and hole states are asymmetric such that
one of the two resulting particles tends to gain more energy than the other.
To contribute to photocurrent generation, hot free
holes need to overcome the transport barrier due to hBN band gap. 
The excess energy required is estimated to be 0.4 eV or
greater (Figure \ref{f1}b) \cite{NanoLett2017vu}. Our DFT modeling demonstrates that
holes derived from recombination of both A and B excitons
possess sufficient excess kinetic energies ($>$ 1 eV) to contribute to photocurrent (Figure \ref{f2}b).


In summary, we have shown that EEA in 2D semiconductors
can be harnessed in photocarrier transport in vdW
heterostructures. EEA rate in 2D semiconductor is significantly
enhanced compared to bulk semiconductor to generate clear
photoresponse in the excitation power range of ordinary
optoelectronic devices, corresponding to photocarrier generation rates greater than $10^{19}$ cm$^{-2}$s$^{-1}$.
We note that the observed effect is not limited to monolayer MoS$_2$.
We have also observed the equivalent effects in devices based on FLG/hBN/WS$_2$ 
heterostructure despite higher barrier that the holes must overcome to yield photocurrent 
(see Supporting Information S5). Although the efficiency of these unoptimized
devices is limited, we envision that an intelligent heterostructure design by material selection and 
band gap engineering will enable improved energy-harvesting devices by
exploiting EEA processes in 2D semiconductors.

\subsection*{Contributions}

G.E. and E.L. conceived the idea of the work. E.L. fabricated
the MIS heterostructure devices and performed electro-optical
measurements and other sample characterizations with
guidance from I.V. and D.V. I.V. and D.V. devised and
assembled the necessary electro-optical setup. D.Y. theoretically explored the EEA generation channels under supervision
of F.P. and M.T. K.W. and T.T. provided hBN crystals. M.T.
and G.E. managed the project, analyzed the data, devised the
model, and wrote the manuscript with inputs from E.L., D.Y.
and F.P.

\section*{Acknowledgements}

G.E. acknowledge the Singapore National Research Foundation (NRF) for funding the research under the Medium Sized
Centre Programme R-723-000-001-281. G.E. also acknowledges support from the Ministry of Education (MOE),
Singapore, under AcRF Tier 2 (MOE2017T2-1-134) and
Tier 3 (MOE2018-T3-1-005). K.W. and T.T. acknowledge
support from the Elemental Strategy Initiative conducted by
the MEXT, Japan and the CREST (JPMJCR15F3), JST. M.T.
acknowledges the Director's Senior Research Fellowship from
the Centre for Advanced 2D Materials at the National
University of Singapore (funded jointly through Singapore
NRF Medium Sized Centre Programme [R-723-000-001-281]
and NUS Young Investigator Award [R-607-000-236-133]).
D.Y. and F.P. were funded by the Collaborative Research
Center (SFB) 767 of the German Research Foundation
(DFG). An important part of the numerical modeling was
carried out on the computational resources of the bwHPC
program, namely the bwUniCluster and the JUSTUS HPC
facility.

\subsection*{Competing Interests}
The authors declare no competing interests.

\section*{Additional Information}

\textbf{Supporting Information} is available free of charge at
\url{https://pubs.acs.org/doi/10.1021/acs.nanolett.9b04756}

\textbf{Correspondence and requests for supplemental materials} should be addressed to either M.T. 
(theory, c2dmt@nus.edu.sg) or G.E. (experiment, g.eda@nus.edu.sg).\\

\bibliography{XXA.bib}

\end{document}